\pdfoutput=1
\RequirePackage{fix-cm}
\documentclass[twocolumn,epjc3]{svjour3}  
\def\makeheadbox{\relax}
\smartqed  % flush right qed marks, e.g. at end of proof
\RequirePackage{graphicx}
\RequirePackage{xspace}
\RequirePackage{amsmath}
\usepackage{subcaption}
\usepackage{siunitx}
\usepackage{cite}

\newcommand{\pT}{\ensuremath{p_\text{T}}\xspace}

\newcommand{\pt}{\ensuremath{p_{\text{T}}}\xspace}

\newcommand{\met}{\ensuremath{E_{\text{T}}^{\text{miss}}}}

\newcommand{\ninoone}{\ensuremath{\tilde{\chi}_{1}^{0}}\xspace}

\newcommand{\ninotwo}{\ensuremath{\tilde{\chi}_{2}^{0}}\xspace}

\newcommand{\nhits}{\ensuremath{N_{\mathrm{layer}}^{\mathrm{hit}}}\xspace}

\newcommand{\mupu}{\ensuremath{\left<\mu\right>\xspace}}

\def\chinoonepm{\ensuremath{\mathchoice%
      {\displaystyle\raise.4ex\hbox{$\displaystyle\tilde\chi^\pm_1$}}%
         {\textstyle\raise.4ex\hbox{$\textstyle\tilde\chi^\pm_1$}}%
       {\scriptstyle\raise.3ex\hbox{$\scriptstyle\tilde\chi^\pm_1$}}%
 {\scriptscriptstyle\raise.3ex\hbox{$\scriptscriptstyle\tilde\chi^\pm_1$}}}}

\journalname{Eur. Phys. J. C}
\begin{document}

\title{Discovery reach for wino and higgsino dark matter with a disappearing track signature at a 100 TeV $pp$ collider%\thanksref{t1}
}

\author{
   Masahiko Saito\thanksref{addr1}
        \and
   Ryu Sawada\thanksref{e1,addr1}
        \and
   Koji Terashi\thanksref{e2,addr1}
        \and
   Shoji Asai\thanksref{addr1}
}

\thankstext{e1}{e-mail: ryu.sawada@cern.ch}
\thankstext{e2}{e-mail: koji.terashi@cern.ch}

\institute{International Center for Elementary Particle Physics and Department of Physics, The University of Tokyo\label{addr1}}

\maketitle

\begin{abstract}
Within the theory of supersymmetry, the lightest neutralino is a dark matter candidate and is often assumed to be the lightest supersymmetric particle (LSP) as well. 
If the neutral wino or higgsino is dark matter, the upper limit of the LSP mass is determined by the observed relic density of dark matter.
If the LSP is a nearly-pure neutral state of the wino or higgsino, the lightest chargino state is expected to have a significant lifetime
due to a tiny mass difference between the LSP and the chargino.
This article presents discovery potential of the 100~TeV future circular hadron collider (FCC) for the wino and higgsino dark matter 
using a disappearing-track signature. 
The search strategy to extend the discovery reach to the thermal limits of wino/higgsino dark matter is discussed with detailed studies on the background rate and the reference design of the FCC-hadron detector under possible running scenarios of the FCC-hadron machine.
A proposal of  modifying the detector layout and several ideas to improve the sensitivity further are also discussed.
\end{abstract}

\section{Introduction}
\label{sec:intro}
Astrophysical observations of galaxies and the large-scale structure in the universe strongly indicate that dark matter
predominate the matter contents in the universe. The nature of dark matter is, however, still unknown and 
the Standard Model (SM) of particle physics has no counterparticles acting as dark matter.
Within the theory of supersymmetry (SUSY), a dark matter candidate is a neutralino, which is a 
mixed state of neutral supersymmetric partners of the SM U(1)$\times$SU(2) gauge bosons and the two SU(2) Higgs doublets.
The lightest neutralino (\ninoone) is often assumed to be the lightest supersymmetric particle (LSP).
If we assume that dark matter is the neutral wino~(higgsino), which is a supersymmetric partner of the SM SU(2) gauge boson~(Higgs boson) and that it was produced in thermal processes, the upper limit of the LSP mass is determined by the observed relic density of dark matter to be 3 (1)~TeV for the wino (higgsino) LSP scenario~\cite{HISANO200734, CIRELLI2007152}.

Wino LSP is predicted naturally by Anomaly Mediated SUSY Breaking (AMSB) models~\cite{Giudice:1998xp,Randall:1998uk} and Pure Gravity Mediation (PGM) models~\cite{IBE2012374, PhysRevD.85.095011, PhysRevD.87.015028}.
Pure wino dark matter has been excluded by indirect searches~\cite{Cohen_2013, Fan2013, PhysRevD.94.115019} when cusped dark matter profiles in the Milky Way and dwarf galaxies are assumed. However, uncertainties in the dark matter distributions in galaxies are considerably large~\cite{Nesti_2013, Sugai2016}. When a cored profile is assumed, only a narrow range of the wino dark-matter mass around the Sommerfeld resonance at $\sim2.4$~TeV is excluded~\cite{Cohen_2013}.
Direct dark matter search results do not exclude the wino dark matter because of the extremely small direct-detection cross section of the pure wino~\cite{HISANO2010311}.

Very-pure higgsino dark matter has been excluded, in the case of very heavy gauginos, due to the absence of inelastic dark matter scattering in direct detection experiments~\cite{Nagata2015}.
If light gauginos are assumed, the higgsino-only dark matter is also excluded due to large mixing under the requirement of a
small fine tuning~\cite{Baer2018}. In such natural SUSY scenarios, the higgsinos are allowed to make up only a portion of the dark matter~\cite{Baer2018}.
Dark matter scenarios with mixed higgsino/gaugino states (so called ``well-tempered''~\cite{ARKANIHAMED2006108}) are also getting less appealing 
due to direct search bounds~\cite{BADZIAK2017226, Beneke2017}. 
However, if a certain amount of fine tuning is allowed, the neutral higgsinos could have a mass splitting due to non-negligible gaugino mixing, and 
therefore are not excluded by direct searches due to the absence of vector interactions.
In this scenario, the nearly-pure higgsino-only dark matter with masses around the thermal limit, which is a subject of this paper, is not constrained~\cite{Nagata2015, Kowalska2018} by direct searches due to the small neutralino-nucleon cross section or by indirect searches because of the negligible Sommerfeld enhancement in the annihilation cross section.

The mass difference of the neutral wino and the charged winos is dominated by the radiative contributions from SM particles when other SUSY particles are a few times heavier than the winos. Contributions from heavier SUSY particles are negligible~\cite{Ibe:2012sx} due to a suppression by the large masses.
As shown in Ref.~\cite{Ibe:2012sx}, the mass difference is nearly constant (160~MeV) for the wino with masses larger than about $500$~GeV.
The SUSY particle contributions to the mass difference of higgsinos are non-negligible.
However, when gauginos are heavier than $\mathcal{O}$(100)~TeV considered in this paper, 
the mass difference of higgsinos is almost constant ($\sim350$~MeV) because SM contributions dominate~\cite{Nagata2015}.
Due to the small mass difference, the \chinoonepm{} has a long lifetime of approximately 0.2~(0.023)~ns for 
the 3~TeV wino~(1~TeV higgsino) at the thermal limit~\cite{1126-6708-2003-03-045}.

With this mass difference, the \chinoonepm{} decays into a \ninoone{} and a charged pion predominantly. The \ninoone{} passes through the detector without any electromagnetic or strong interactions while the pion is not reconstructed as a track in the inner-tracking detector 
due to very low transverse momentum (\pt). 
Therefore, such a long-lived chargino is observed in the detector as a short `disappearing' track, which has hits in several silicon-detector layers near the interaction point and no associated hits after a certain radius, with a typical length of $O$(1--10)~cm in collider experiments~\cite{PhysRevD.55.330,Chen:1999yf,Asai2008185}.

In this paper, a search strategy for the long-lived charginos with a disappearing-track signature at the 100~TeV FCC-hadron machine (FCC-hh) is discussed with the reference FCC-hadron detector. The expected sensitivity is presented under the scenario of 30~ab$^{-1}$ of $pp$ data collected under possible running conditions of the FCC-hh.
In earlier studies on the prospects of disappearing-track searches in future $pp$ colliders~\cite{Low2014, Cirelli2015, Mahbubani2017, FUKUDA2018306, PhysRevD.98.035026, Cao2018}, effects of multiple $pp$ collisions occurring simultaneously with a signal event (pileup) were assumed not to alter the results significantly by improving the analysis method or the background rate was scaled with a certain factor without detailed studies. In this paper, the background rates were estimated based on the detector layout and simulation.
In addition, several ideas that could potentially improve the sensitivity further are discussed.

\section{Simulation samples}
\label{sec:simulation}
Simulated samples of Monte Carlo~(MC) events are used to obtain the kinematic distributions of the signal and background processes.
The wino mass spectrum is calculated using \textsc{Softsusy}~3.7.3~\cite{Allanach:2001kg} assuming the minimum Anomaly Mediated SUSY Breaking~(AMSB) model with tan$\beta=5$, positive sign of the higgsino-mass term and the universal scalar mass~($m_0$) of 20~TeV. This setting naturally accommodates the observed Higgs boson mass of 125~GeV without maximal mixing.
The higgsino mass spectrum is calculated by \textsc{Susyhit}~\cite{Djouadi:2006bz} assuming the general minimal 
supersymmetric extension of the SM (MSSM) with tan$\beta=5$. Other SUSY particle masses are set to 100~TeV to prevent the higgsino from being affected from loop effects.
Only the pure wino and higgsino LSP scenarios are considered here.
The production cross sections for the winos and higgsinos in 100~TeV $pp$ collisions are calculated using \textsc{Prospino2}~\cite{Beenakker:1996ch}.
The kinematic distributions are simulated using \textsc{MadGraph5\_aMC@NLO}~2.3.3~\cite{Alwall:2014} assuming the wino spectra.
The same kinematic distributions are assumed for the wino and higgsino if they have the same masses. 
In this study, pair-production processes of charginos and neutralinos are considered:
$\tilde{\chi}_{1}^{\pm}\tilde{\chi}_{1}^{\mp}$ and $\tilde{\chi}_{1}^{\pm}\tilde{\chi}_{1}^0$ for the wino scenario, and 
$\tilde{\chi}_{1}^{\pm}\tilde{\chi}_{1}^{\mp}$,  $\tilde{\chi}_{1}^{\pm}\tilde{\chi}_{1}^0$ and $\tilde{\chi}_{1}^{\pm}\tilde{\chi}_{2}^0$ for the higgsino scenario. 
As the \ninoone{} and \ninotwo{} are degenerate in the higgsino case, the difference in the cross section or decay processes between
$\tilde{\chi}_{1}^{\pm}\tilde{\chi}_{1}^0$ and $\tilde{\chi}_{1}^{\pm}\tilde{\chi}_{2}^0$ is negligible.
Therefore, only the $\tilde{\chi}_{1}^{\pm}\tilde{\chi}_{1}^0$ process is generated and the cross section is doubled to account 
for $\tilde{\chi}_{1}^{\pm}\tilde{\chi}_{2}^0$.
Minimum-bias collisions are produced using \textsc{Pythia}~8.230~\cite{Sjostrand:2014zea} and are overlaid to evaluate contribution from non-genuine tracks caused by the random combination of hits in the inner-tracking detector, which are referred to as fake tracks hereafter. The details of the minimum-bias event overlay are described in Sect.~\ref{sec:background}.

SM background processes such as $W/Z$-boson production process in association with jets, top-quark pair and single top-quark production processes
are simulated using \textsc{MadGraph5\_aMC@NLO} and the generated samples are used to evaluate the selection efficiency.
The kinematic distributions of jets, electrons, muons and the magnitude of missing transverse momentum (\met) are processed using \textsc{Delphes}~3.4.2~\cite{deFavereau2014} with the reference FCC-hh run card to account for the detector reconstruction efficiency and energy/momentum measurement. Jets are reconstructed from particle-flow candidates using the anti-$k_t$ algorithm~\cite{Cacciari:2008gp} with a distance parameter of 0.4.

\section{Inner Tracker Layout}
\label{sec:layout}
The reconstruction of short tracks relies on the performance of the inner-most tracking system. This paper considers the reference design of the FCC-hadron detector~\cite{FCC-hh_detector} as a baseline. For the inner tracker, the most relevant part for the disappearing-track search is the inner-pixel layers, in particular those in the barrel region where the signal
acceptance is high. In the present study three different layouts are considered for the barrel region 
to evaluate impact of the inner-tracker configuration.

\begin{figure}[htbp]
  \centering
  \begin{subfigure}{1.0\columnwidth}
    \centering
    \includegraphics[width=\columnwidth]{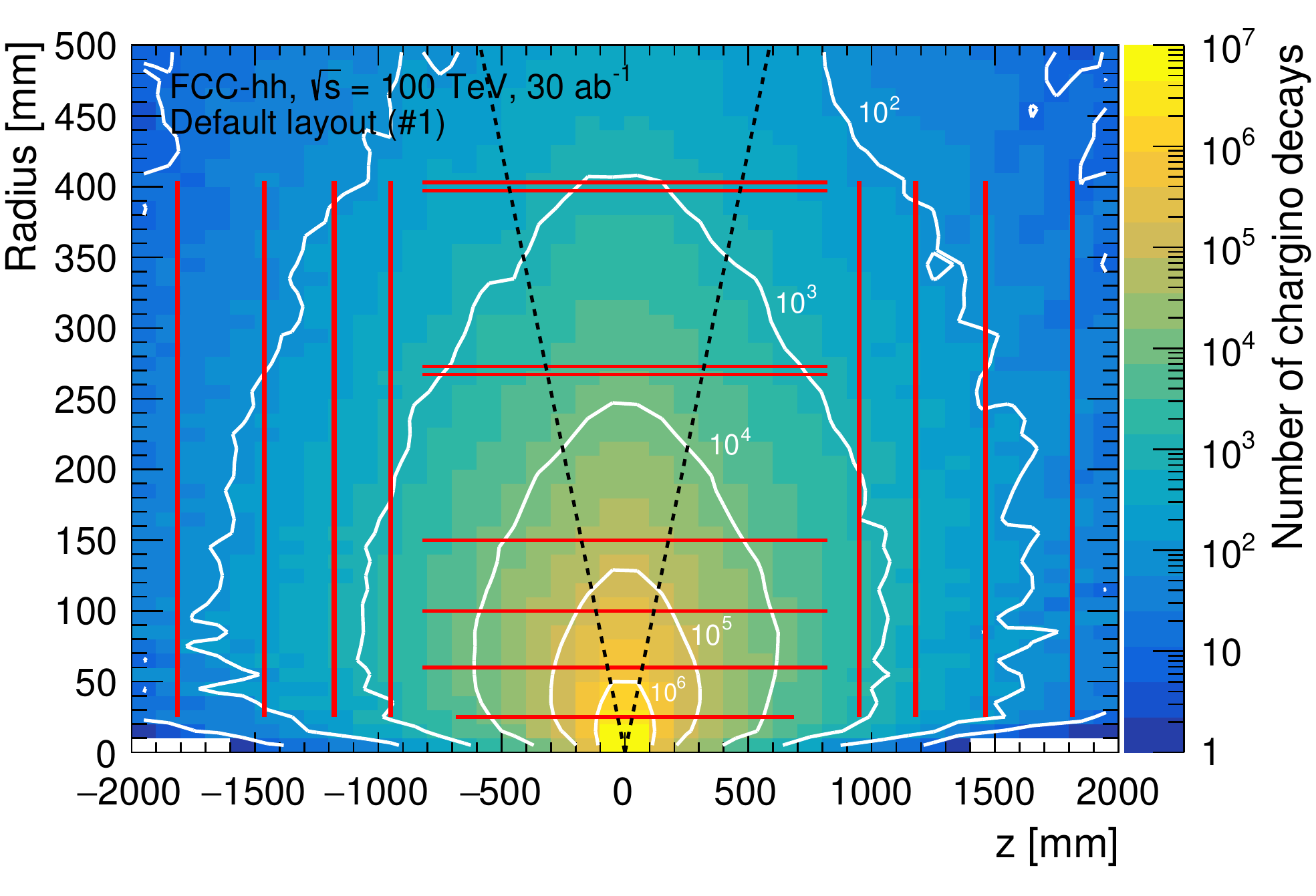}
  \end{subfigure}
  \begin{subfigure}{1.0\columnwidth}
    \centering
    \includegraphics[width=\columnwidth]{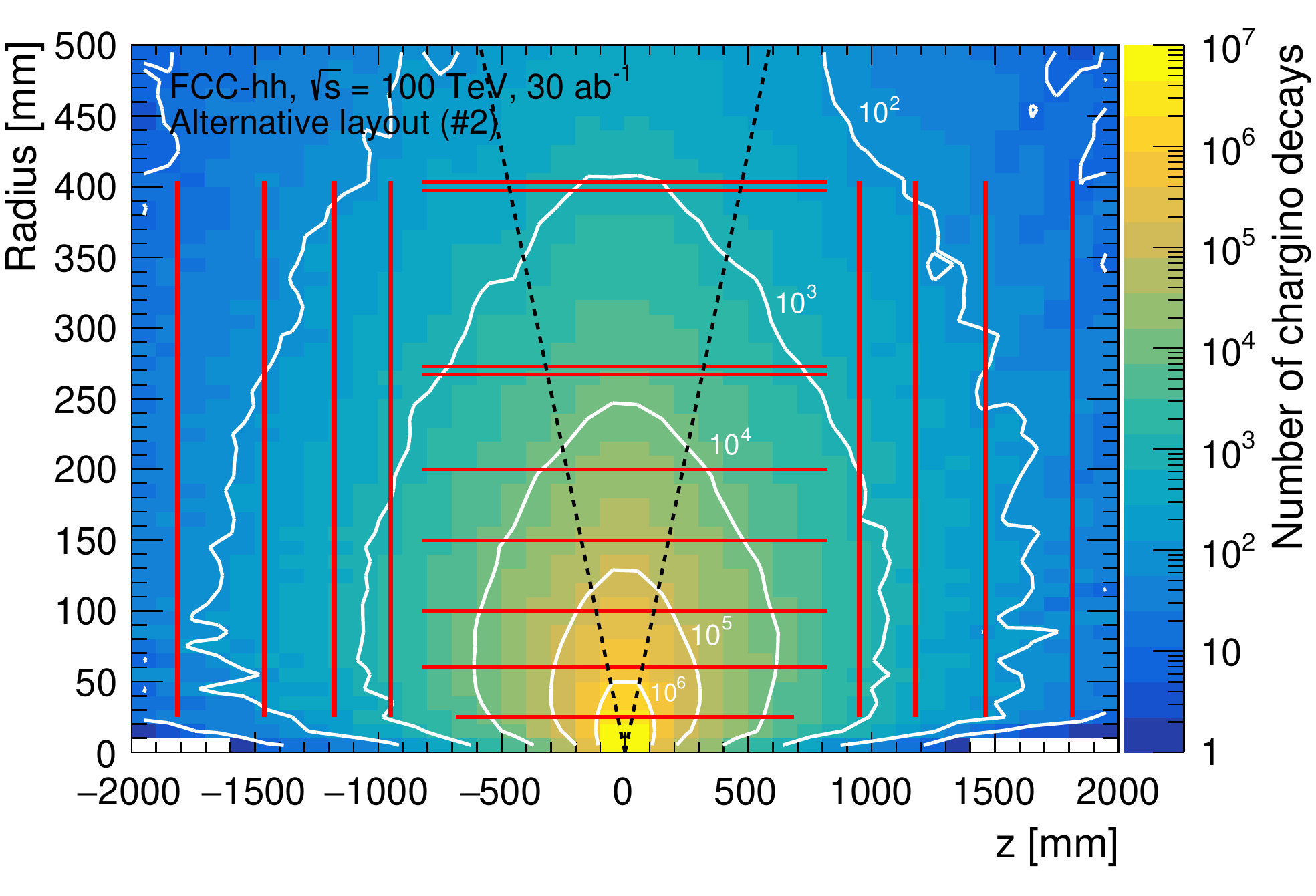}
  \end{subfigure}
  \begin{subfigure}{1.0\columnwidth}
    \centering
    \includegraphics[width=\columnwidth]{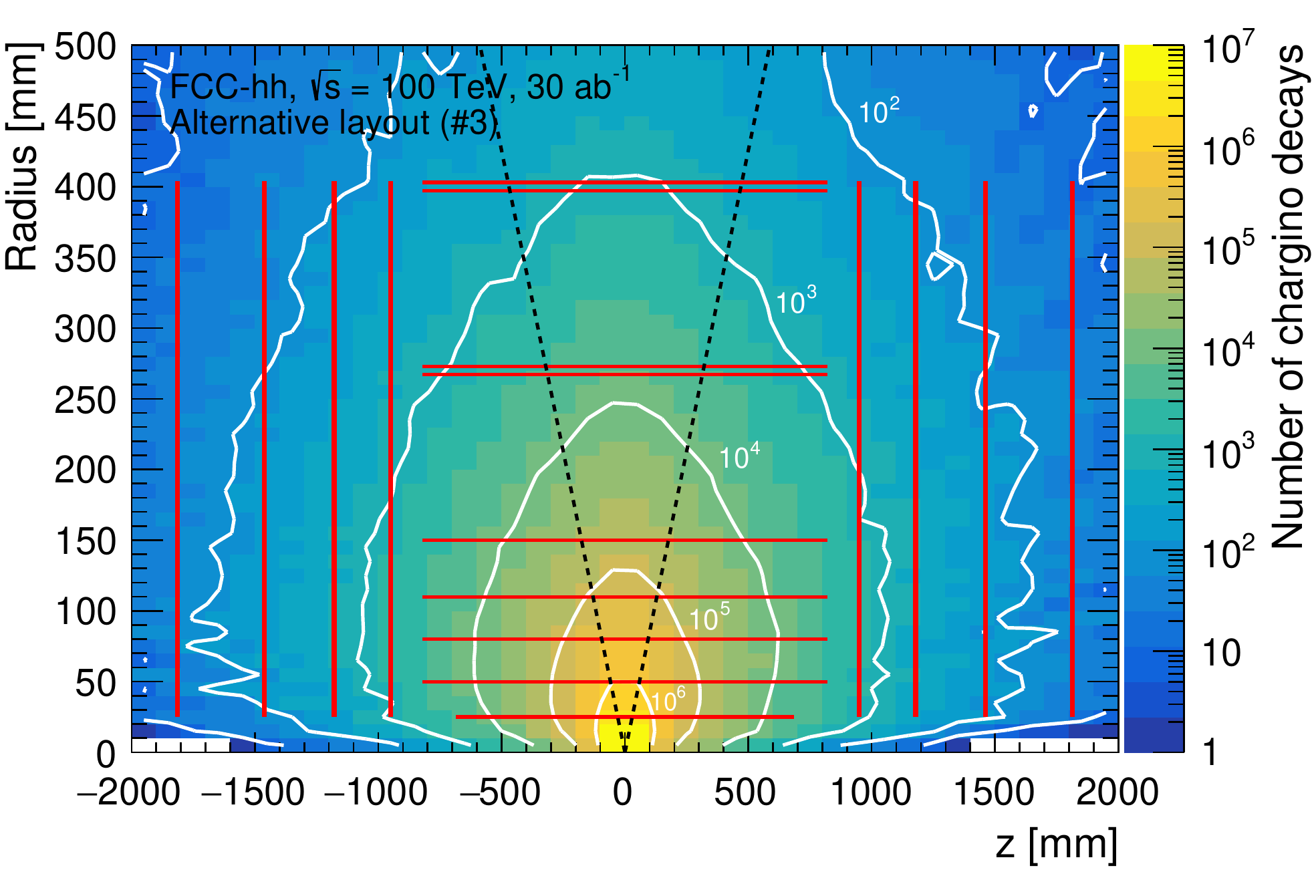}
  \end{subfigure}
  \caption{
    Three barrel inner-tracker layouts considered in this study, the default layout~\#1 (top), the alternative layout~\#2 (middle) and the alternative layout~\#3 (bottom). The difference between different layouts is restricted 
    within $R\leq200$~mm and $|z|\leq82$~cm, and the detector configuration is identical outside the region. 
    The contour drawn behind the layouts shows the number of chargino decays in 3~TeV wino signal events 
    with 30~ab$^{-1}$ at a given position. The analysis considers the region $|\eta|<1$, denoted by the dotted lines.
    }
  \label{figure:DecayPositionWithLayout}
\end{figure}

The default layout (labelled \#1) is the one in the reference design, containing four pixel layers within $R=150$~mm in radial distance from the beam line. In the second layout (\#2) a pixel layer is added at $R=200$~mm to the default layout and all the other layers are unchanged. The third layout (\#3) considers an additional pixel layer at $R=150$~mm and the inner four pixel layers are moved closer to the beam line. The $R$-$z$ views of all the three layouts are shown in Fig.~\ref{figure:DecayPositionWithLayout}.\footnote{
We use a right-handed coordinate system, with the $x$-, $y$-, and $z$-axis are defined by the horizontal, the vertical and the beam directions respectively. Its origin is at the nominal $pp$ interaction point.
The pseudorapidity $\eta$ is defined in terms of the polar angle $\theta$ by $\eta=-\ln\tan(\theta/2)$.
}
The outer macro-pixel layers located at $R\geq270$~mm in the barrel region or the forward tracking system are unchanged from the reference design for all the three layouts. The first five inner layers are most relevant for the short-track reconstruction as described below. Their radial positions considered in this study are summarized in Table~\ref{table:LayerPosition}.

\begin{table}[t]
  \centering
  \caption{Radial distances (in mm) of the first five layers from the beam line for the three inner-tracker layouts considered in this study.}
  \label{table:LayerPosition}
  \begin{tabular}{lccccc}
    \hline\noalign{\smallskip}
    Layer \#    & 1 & 2 & 3 & 4 & 5 \\
    \noalign{\smallskip}\hline\noalign{\smallskip}
    Default layout~(\#1)      & 25 & 60 & 100 & 150 & 270    \\
    Alternative layout~(\#2) &  25 & 60 & 100 & 150 & 200   \\
    Alternative layout~(\#3) &  25 & 50 & 80 & 110 & 150   \\
    \noalign{\smallskip}\hline
  \end{tabular}
\end{table}

\section{Event Selection}
\label{sec:selection}
\begin{figure}[t]
  \centering
  \begin{subfigure}{1.0\columnwidth}
    \centering
    \includegraphics[width=\columnwidth]{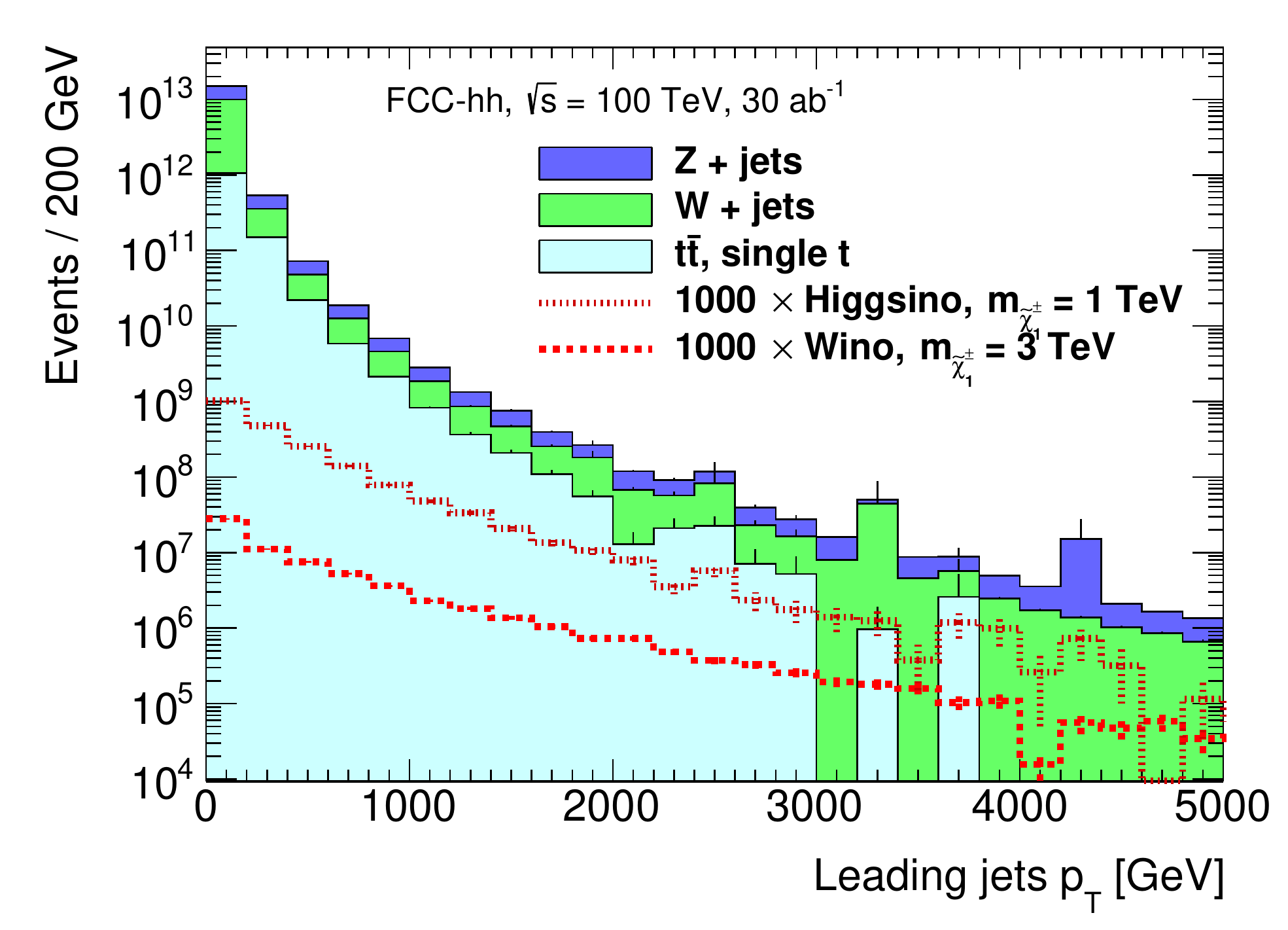}
  \end{subfigure}
  \begin{subfigure}{1.0\columnwidth}
    \centering
    \includegraphics[width=\columnwidth]{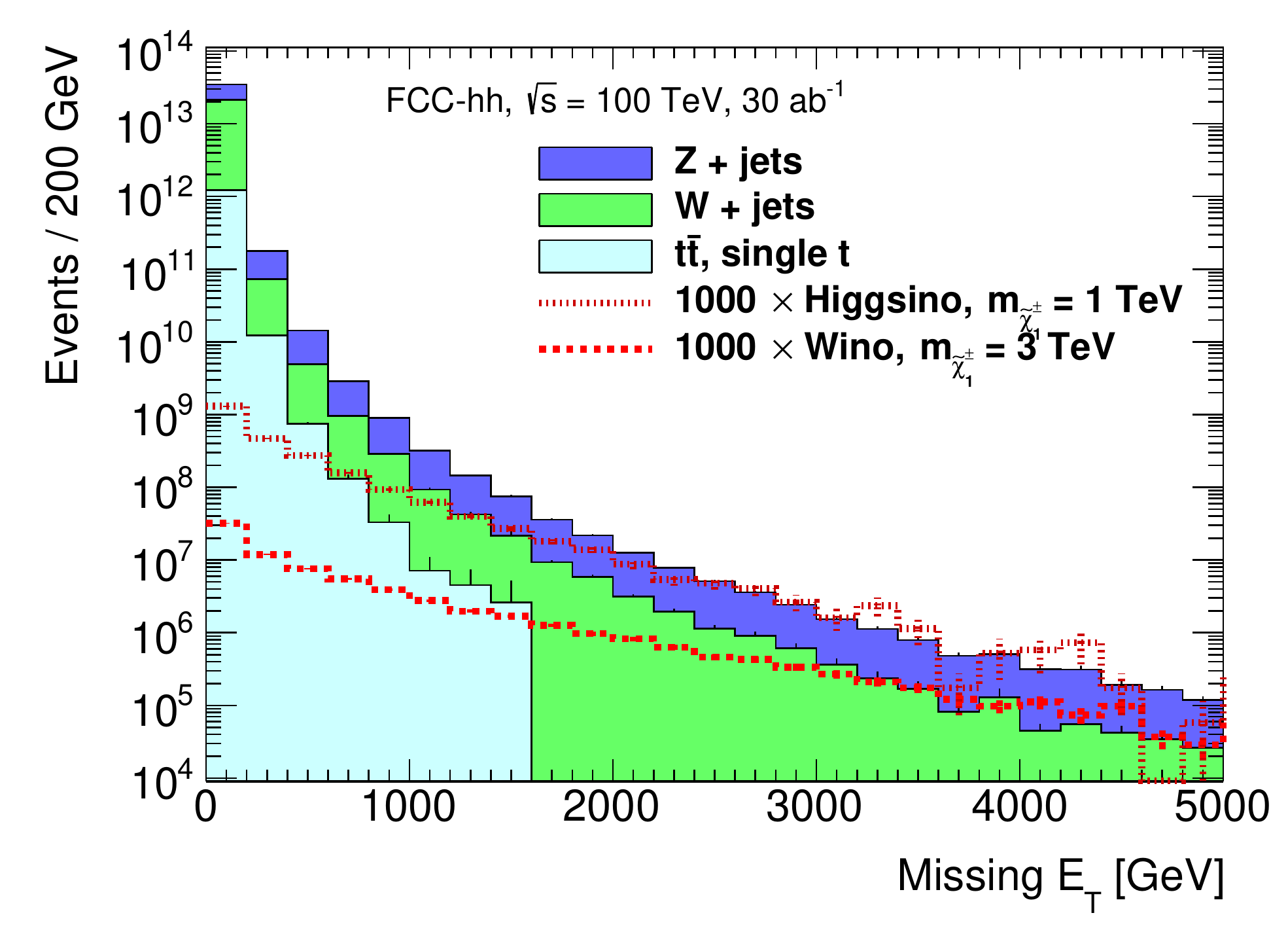}
  \end{subfigure}
  \caption{Leading jet \pt{} (top) and \met{} (bottom) distributions after removing events containing isolated leptons with 30~ab$^{-1}$ at $\sqrt{s}=100$~TeV. 
  The SM backgrounds from $W/Z$+jets and top production processes are shown as filled histograms. 
  Also shown as dashed (dotted) line is the 3 (1)~TeV wino (higgsino) signal scaled up by a factor 1000.
          }
  \label{figure:JetMETPt}
\end{figure}

The chargino/neutralino-pair production is characterized by the presence of significant \met{} associated with the \ninoone{} and one or two disappearing tracks from the \chinoonepm. The SM background is suppressed by requiring an existence of a high-$\pt${} disappearing track. The signal event will presumably contain high-\pt{} jet(s) as well, originating from initial state radiation, and they provide the means of triggering on the event in conjunction with \met.
The analysis therefore selects events which contain at least one high-\pt{} jets, large \met{} and 
no isolated lepton (electron or muon) with $\pT>10$~GeV. 
Figure~\ref{figure:JetMETPt} shows the leading jet \pt{} and \met{} distributions of signal and background events selected with the isolated lepton veto and no requirement on the jet \pt{} or \met. The thresholds on the jet \pt{} and \met{} are determined, for each configuration of the inner-tracker layouts, by maximizing the sensitivity based on the signal acceptance and background rate
(summarized later in Tables~\ref{table:ResultSensitivitymu200} and \ref{table:ResultSensitivitymu500}).
Although the background yield can be reduced significantly by applying high thresholds to the leading jet \pt{} and \met, there will be no discovery power for the wino or higgsino at the thermal limit if the disappearing-track information is not used in the search.

\begin{figure}[t]
  \centering
  \includegraphics[width=1.0\columnwidth]{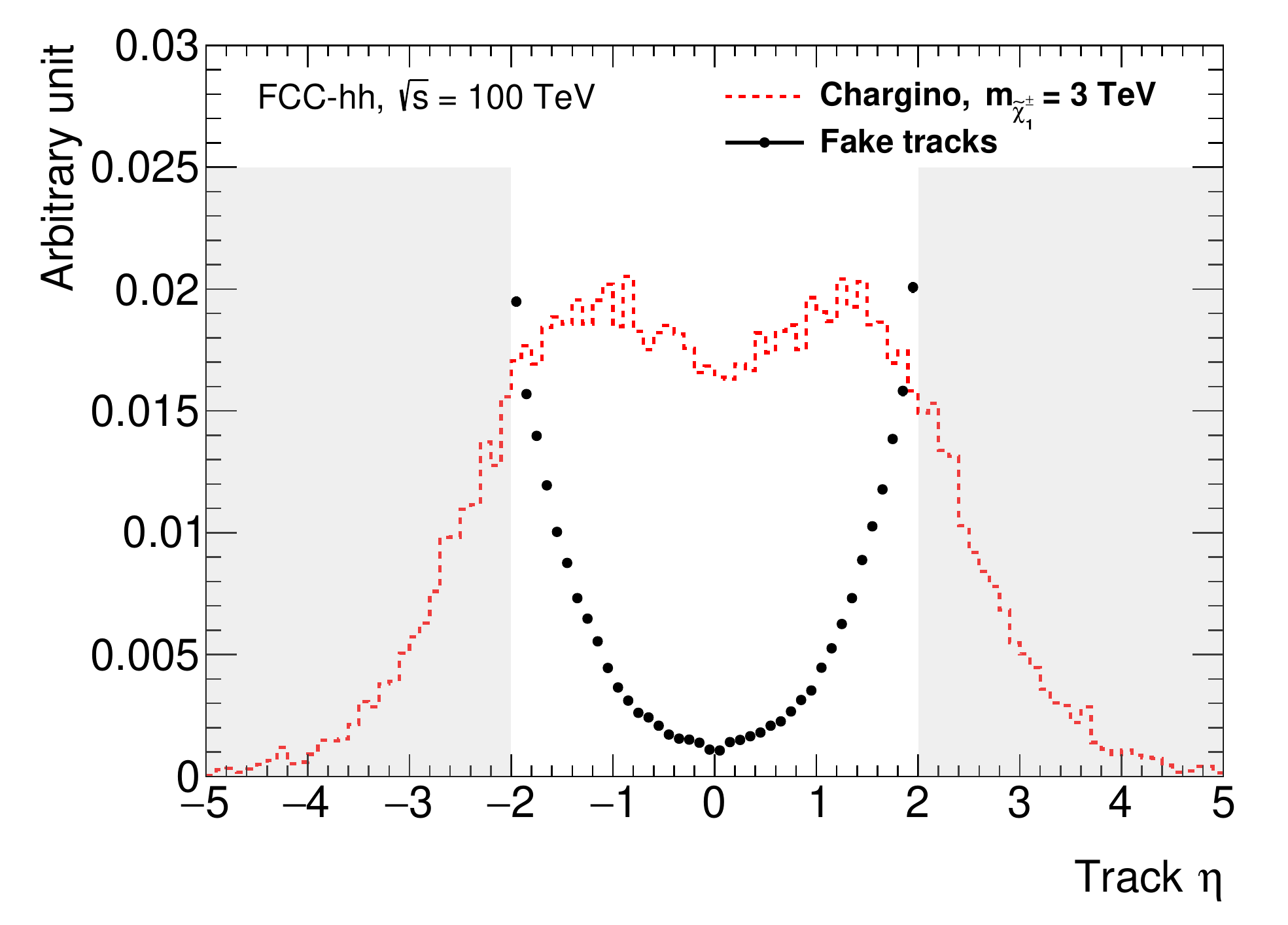}
  \caption{$\eta$ distributions of fake tracks in minimum-bias collision events and chargino tracks in 3~TeV wino signal events. The normalization is arbitrary. The details of fake track reconstruction is described in Sect.~\ref{subsec:fake_background}. The fake-track distribution is shown only within the range $|\eta|<2$.}
  \label{figure:fakeEtaDist}
\end{figure}

The chargino-track reconstruction requires the particle to traverse a certain minimum number of inner-tracking layers before decay. The ATLAS search with 36~fb$^{-1}$ of Run 2 data at $\sqrt{s}=13$~TeV~\cite{Aaboud:2017mpt} used a short-track reconstruction with only hits in the four pixel layers. As the chargino lifetime becomes short, the acceptance for the chargino-track reconstruction increases with decreasing radial position of the outermost layer necessary for the track reconstruction. On the other hand, the use of short tracks with less number of layer hits will usually increase fake tracks (discussed in Sect.~\ref{sec:background}). In order to assess the impact on the sensitivity from different requirements on the number of hits used in the track reconstruction (\nhits), the track reconstruction is performed with the requirements of $\nhits=4$ and $\nhits=5$ separately. When the $\nhits=4$ (5), the four (five) inner-most layers with hits are used to reconstruct tracks and the fifth (sixth) layer or above is not considered in the track reconstruction. 
The contribution from fake tracks increases significantly at high $|\eta|$ due to higher hit occupancy. 
Figure~\ref{figure:fakeEtaDist} shows normalized $\eta$ distributions of fake tracks in minimum-bias collision events (details in Sect.~\ref{subsec:fake_background}) and chargino tracks in 3~TeV wino signal events. Given the dramatic increase of fake background with increasing $|\eta|$, the analysis requires a candidate track to be within $|\eta|<1$ 
(further discussion is in Sect.~\ref{sec:conclusion}).

The signal acceptance for the track requirement is estimated based on the tracker geometry and the chargino lifetime by assuming that the chargino tracks can be reconstructed at 100\% of time if the charginos traverse at least four or five inner-most layers before decay.
Table~\ref{table:SignalAcceptanceResult} summarizes the signal acceptances for the wino and higgsino under the three scenarios of the inner-tracker layouts.

\begin{table*}\sidecaption
  \centering
  \begin{tabular}{cS[round-mode=off]S[round-mode=off]S[round-mode=off]}
    \hline\noalign{\smallskip}
    Layout        & {Default~(\#1)} & {Alternative~(\#2)} & {Alternative~(\#3)} \\
    \noalign{\smallskip}\hline\noalign{\smallskip}
                  & \multicolumn{3}{c}{wino ($m_{\tilde{\chi}_{1}^{\pm}}$ = 3~TeV)} \\
    \noalign{\smallskip}\hline\noalign{\smallskip}
    $\nhits=4$ & 2.5\%     & 2.5\%       & 4.4\%        \\
    $\nhits=5$ & 0.57\%    & 1.3\%       & 2.5\%        \\
    \noalign{\smallskip}\hline\hline\noalign{\smallskip}
                  & \multicolumn{3}{c}{higgsino ($m_{\tilde{\chi}_{1}^{\pm}}$ = 1~TeV)} \\
    \noalign{\smallskip}\hline\noalign{\smallskip}
    $\nhits=4$ & 4.3$\times10^{-3}$\% & 4.3$\times10^{-3}$\% & 1.6$\times10^{-2}$\% \\
    $\nhits=5$ & 2.2$\times10^{-4}$\% & 1.1$\times10^{-3}$\% & 4.3$\times10^{-3}$\% \\
    \noalign{\smallskip}\hline
  \end{tabular}
  \caption{Signal acceptance for the 3~TeV wino and 1~TeV higgsino with $|\eta|<1$ for the three inner-tracker layouts. The acceptances are provided separately for the requirements of $\nhits=4$ and 5. The alternative layout \#3 has significantly higher acceptance than the others because the relevant layers are located closer to the beam line, particularly for the higgsino case with $\nhits=5$.
          }
  \label{table:SignalAcceptanceResult}
\end{table*}

\section{Background estimation}
\label{sec:background}
Backgrounds for the disappearing-track signature are categorized into two components~\cite{Aaboud:2017mpt}.
The first one is physical background, which arises from charged particles (mainly electrons or charged pions) scattered by the material in the inner tracker.
The second one is unphysical background, which arises from fake tracks. The estimate of both background components is discussed separately below.

\subsection{Physical background}
\label{subsec:physbg}
The physical background due to particle scattering is estimated by assuming that it is solely determined by the amount of detector material passed through by particles. Under this assumption the physical background rate at the FCC-hh is obtained by scaling the background rate measured in ATLAS~\cite{Aaboud:2017mpt} by the ratio of the material budgets in the FCC-hh and ATLAS inner trackers~\cite{ATLASMaterialBudget}. 
A support structure of the FCC-hh tracker has not been fully defined yet, so the expected impact from particle scattering with support material is accounted for by assuming the same relative amounts of tracker layers and support material measured in ATLAS. 
The physical background is dominated by $W(\rightarrow \ell\nu)$+jets processes, which have large \met{} and a high-\pt{} track associated with the lepton from the $W$ boson decay.
Therefore, the final estimate of the physical background is obtained by correcting the scaled background rate above for the cross section and kinematic selection efficiency for $W$+jets processes at $\sqrt{s}=100$~TeV.

\subsection{Background with fake tracks}
\label{subsec:fake_background}
The fake-track background is evaluated by counting the number of reconstructed tracks with good quality using simulated minimum-bias events.
The samples of minimum-bias events are generated with the average number of $pp$ interactions per bunch crossing (\mupu) of 200 or 500, and simulated for the FCC-hadron detector using \textsc{Geant4}~\cite{Agostinelli:2002hh}. The two \mupu{} values are assumed to be representative of FCC-pileup conditions.
Pileup collisions are produced with two different models of soft QCD processes: the first model considers only non-diffractive processes while the second one the mixture of diffractive and non-diffractive processes with the relative fraction determined according to the \textsc{Pythia} cross sections\footnote{The non-diffractive processes are set using the {\tt SoftQCD:nonDiffractive} option while the diffractive ones using {\tt SoftQCD:singleDiffractive} and {\tt SoftQCD:doubleDiffractive} options in  \textsc{Pythia}.}.
The former model predicts more fake tracks than the latter due to the absence of diffractive events, and is used as
the nominal pileup model to be more conservative.

\begin{figure}[t]
  \centering
  \includegraphics[width=1.0\columnwidth]{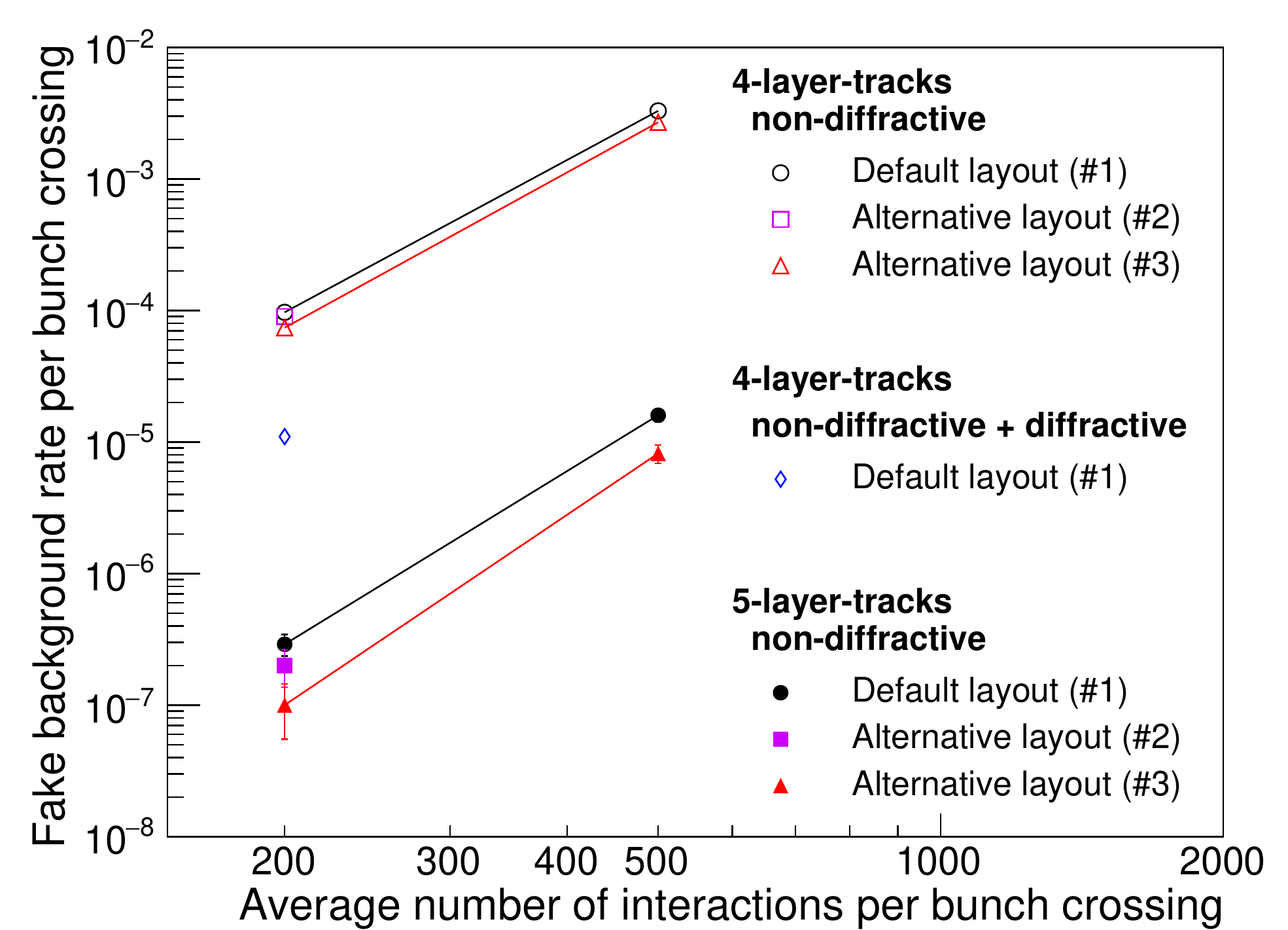}
  \caption{Probability of finding a fake track in an event as a function of the average number of $pp$ interactions per bunch crossing (\mupu). The probabilities for the tracks reconstructed with 
  $\nhits=4$ (5) are shown by the open (filled) markers, and they are presented separately for the 
  three tracker layouts: the default layout (\#1) in black circles, the alternative layouts of \#2 in magenta 
  squares and \#3 in red upward triangles. The probabilities for the layout \#2 are shown only at 
  $\mupu=200$.
  The estimate from the mixed sample of non-diffractive and diffractive events for the default layout is 
  shown by the blue diamond for comparison purpose. 
  }
  \label{figure:fakeRatePileupDeptmp}
\end{figure}

The track reconstruction is performed as follows.
First, the track-seed finding is done by finding a group of inner-tracker hits within a narrow region with a size of about $10^{-3}\times10^{-3}$ in $\eta$ and $\phi$. 
Second, a track is reconstructed by fitting a line with a given curvature to the collection of the hits. The fit is iterated by removing hits which are far from the line.
The following quality criteria are required to select chargino-track candidates: \pt{} $>100$~GeV, $|\eta|< 1$, $\chi^{2}\text{-probability of the track fit}>0.1$, at least one hit in each layer and the association to the hard-scattering vertex by applying the nominal impact-parameter requirement (transverse and longitudinal impact parameters of $|d_{0}|<0.05$~mm and $|z_{0}|<0.5$~mm). 
The final number of fake tracks is obtained by counting the number of tracks with loose impact-parameter requirements ($|d_{0}|<1$~mm, $|z_{0}|<250$~mm) and scaling that to the nominal impact-parameter requirement assuming a uniform distribution of $d_{0}$ and $z_{0}$.
Figure~\ref{figure:fakeRatePileupDeptmp} shows the estimated probability of finding a fake track per event
as a function of the \mupu.
The fake-track finding probability increases significantly with increasing \mupu. However, the probability can be reduced by 2--3 orders of magnitude by changing the \nhits{} requirement from 4 to 5.
The number of fake-track backgrounds is estimated by multiplying the number of SM background events satisfying the kinematic selection criteria (in Sect.~\ref{sec:selection}) by these probabilities.

The time resolution of pixel detectors can be better than 50~ps by utilizing, for example, low-gain avalanche detectors~\cite{Pellegrini2014, Lange2017}.
Therefore, the time information could be used in the track fit as an additional track parameter to potentially improve 
the quality of the track.
The reduction of fake-track background by requiring a good time quality (i.e, small $\chi^2$ in the estimation of track time obtained 
from the associated pixel hits) is investigated using the simulated minimum-bias samples with $\mupu=500$.
In the simulation, the $x$, $y$, $z$ positions and the time of $pp$ collisions are distributed randomly according to Gaussian probability-density functions with the standard deviations of 0.5~mm, 0.5~mm, 50~mm and 160~ps, respectively (without 
assuming any correlations among them). The $\chi^2$ is computed assuming the constant time resolution of 50~ps for a single 
layer hit. The fake-track background is found to be reduced by 96\% by requiring the $\chi^2$-probability to be larger than 0.05 for tracks reconstructed with $\nhits=4$ when $pp$ collisions occur in bunch crossings separated by 25~ns.
The signal efficiency for this selection is approximately 95\%, evaluated using a sample of single muon events.

\section{Result}
\label{sec:result}
The discovery sensitivity with the requirement of $\nhits=5$ is higher than that with $\nhits=4$ due to much lower 5-layer-hit fake-track background, as indicated in Fig.~\ref{figure:fakeRatePileupDeptmp}. The fake-track background rate appears to be rather insensitive to the inner-tracker layout for a fixed \nhits{} requirement. Therefore, the alternative layout \#2 is expected to have sensitivity in-between the default and alternative layout \#3 (hence not shown in the rest of the figures) because the signal acceptance is a factor 2--4 smaller for the \#2 than \#3 (see Table~\ref{table:SignalAcceptanceResult}).

\begin{figure}[t]
  \centering
  \begin{subfigure}{1.0\columnwidth}
    \centering
    \includegraphics[width=\columnwidth]{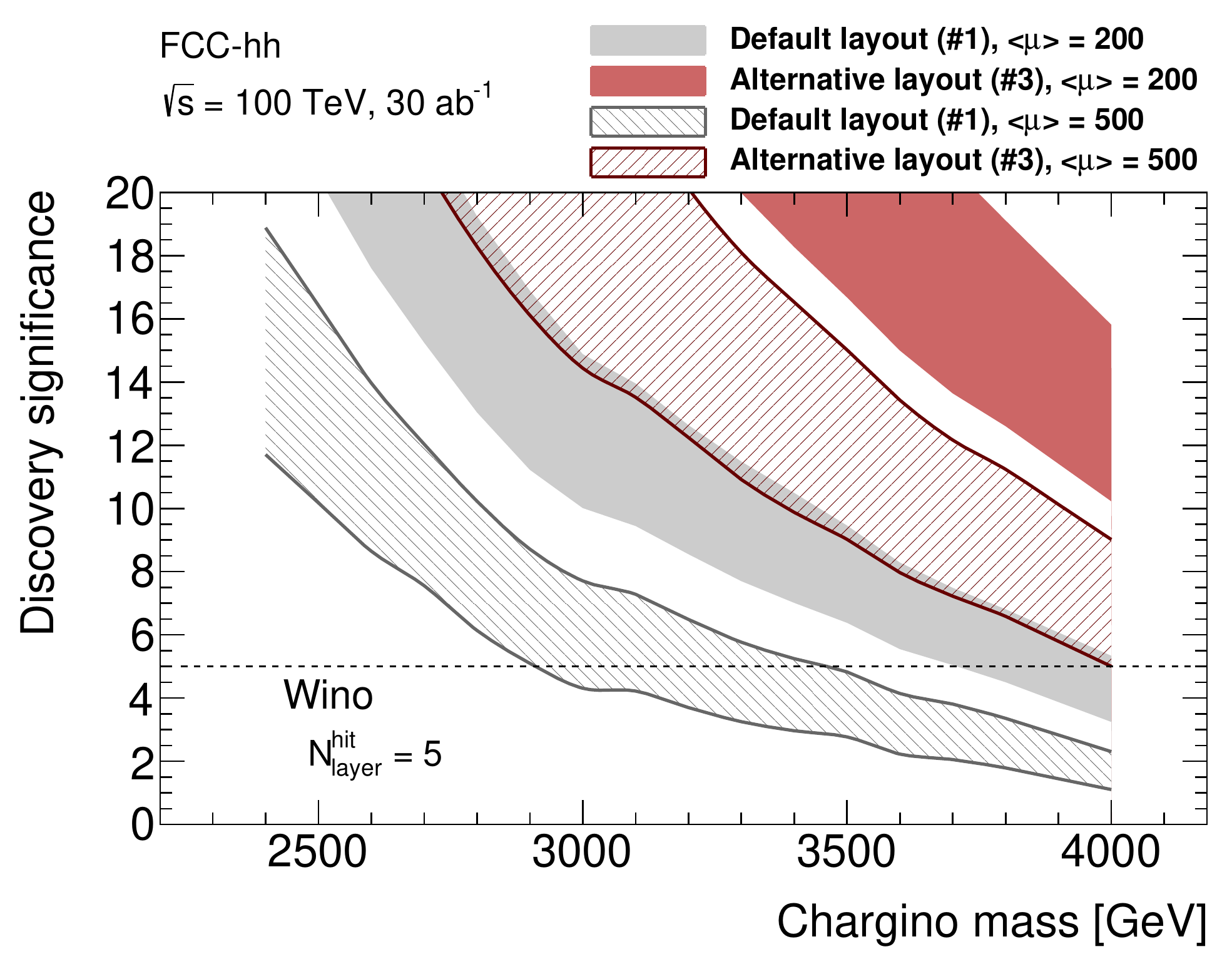}
  \end{subfigure}
  \begin{subfigure}{1.0\columnwidth}
    \centering
    \includegraphics[width=\columnwidth]{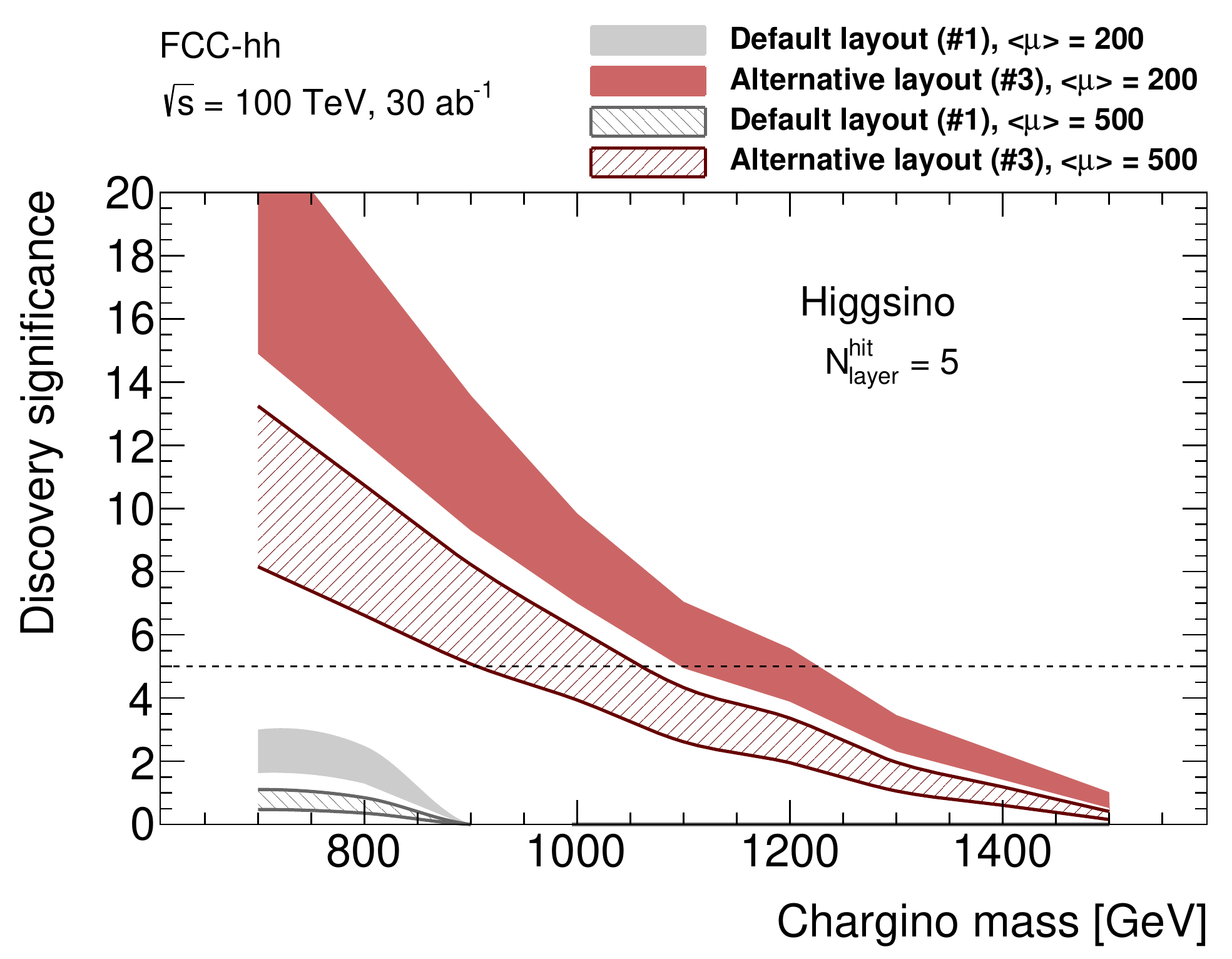}
  \end{subfigure}
  \caption{Expected discovery significance for the wino (top) and higgsino (bottom) models with 30 ab$^{-1}$ 
  with the requirement of $\nhits=5$.
  The grey (red) bands show the significance using the default (alternative) layout \#1 (\#3). 
  The difference between the solid and hatched bands corresponds to the different pileup conditions of $\mupu=200$ and 500. 
  The band width corresponds to the significance variation due to the two models assumed for soft QCD processes.}
  \label{figure:Significance}
\end{figure}

\begin{figure}
  \centering
  \begin{subfigure}{1.0\columnwidth}
    \centering
    \includegraphics[width=\columnwidth]{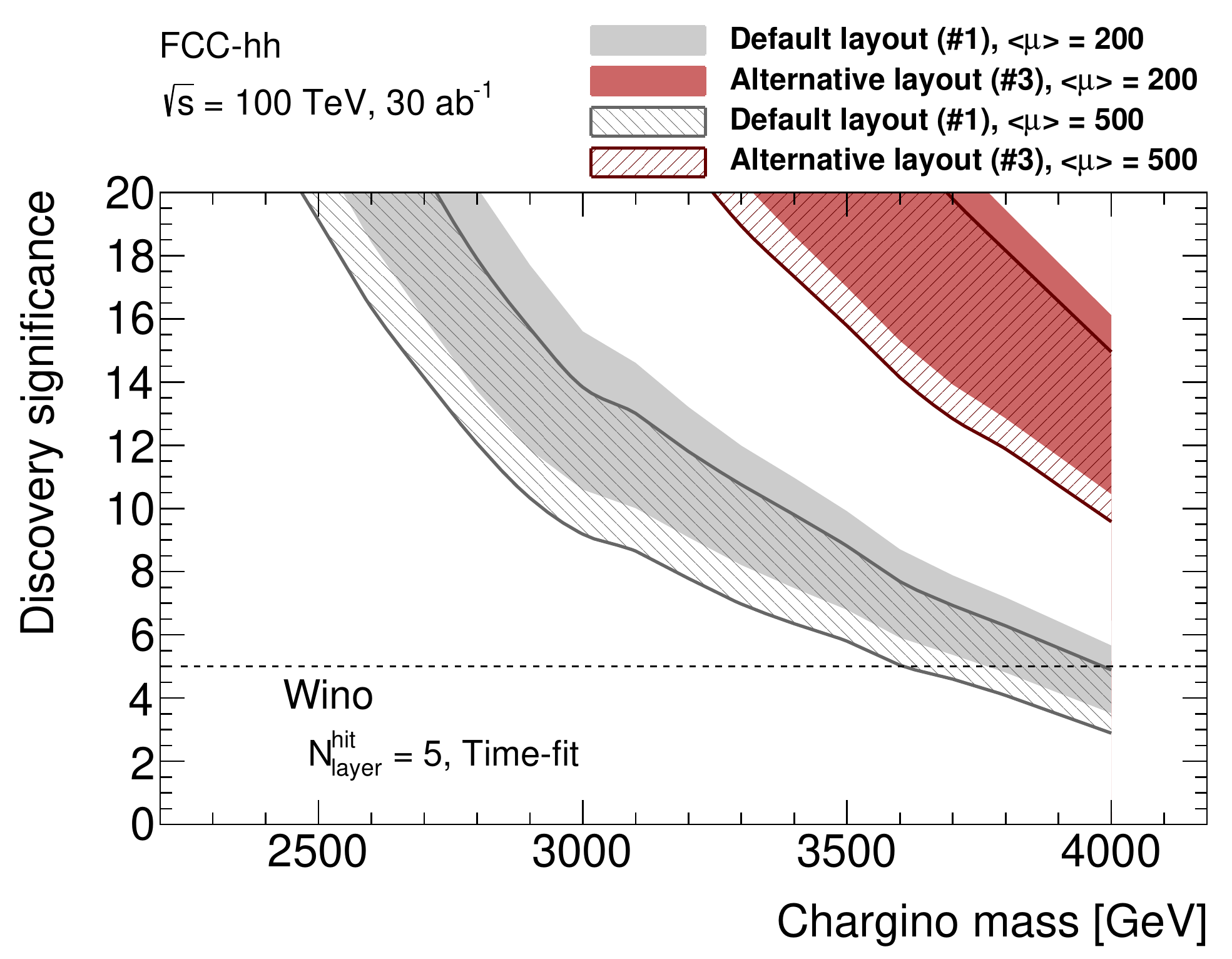}
  \end{subfigure}
  \begin{subfigure}{1.0\columnwidth}
    \centering
    \includegraphics[width=\columnwidth]{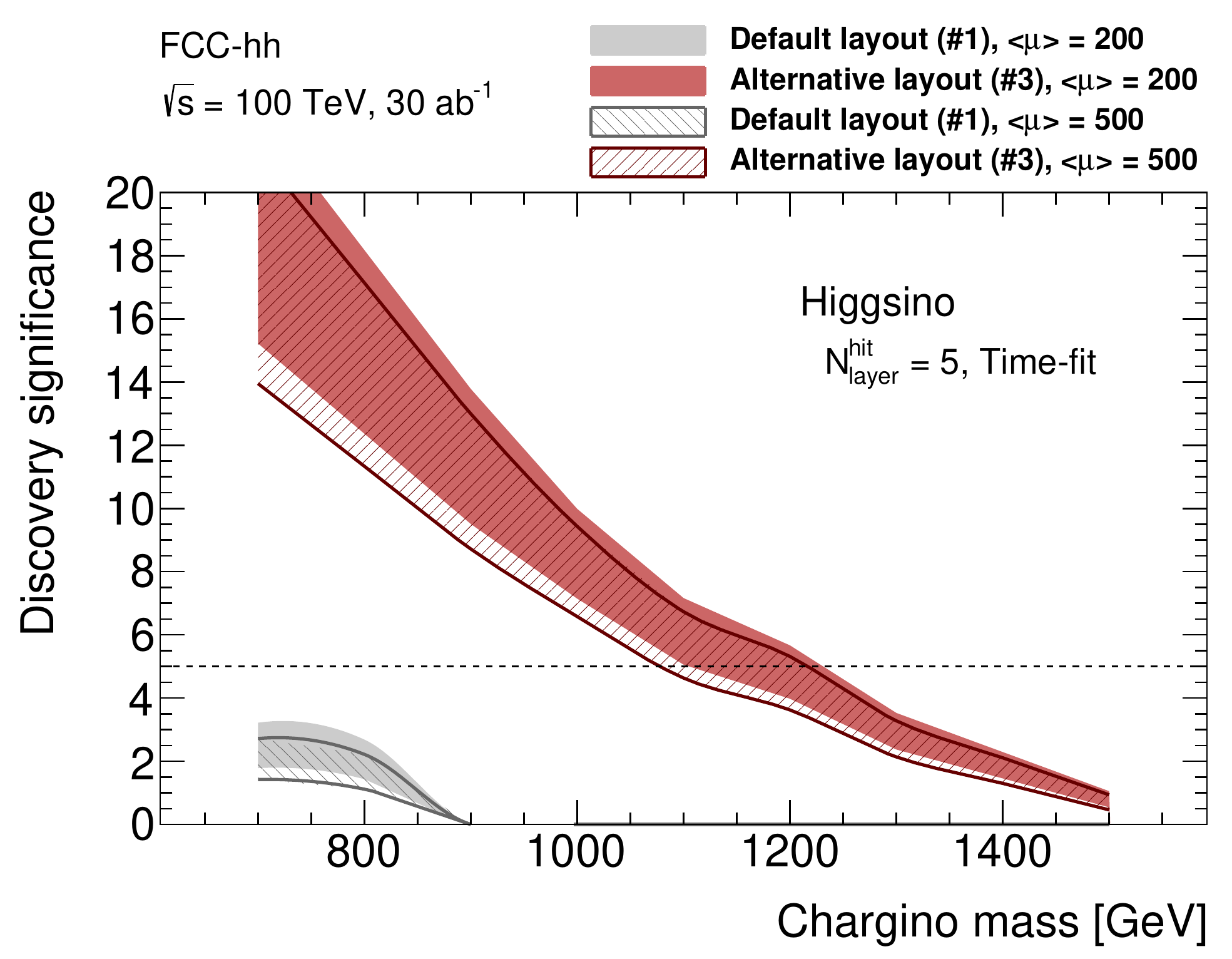}
  \end{subfigure}
  \caption{ Expected discovery significance for the wino (top) and higgsino (bottom) models 
  with 30 ab$^{-1}$ with the requirements of $\nhits=5$ and a good time-fit quality. 
  The background reduction rate with the time information is assumed to be the same for both pileup conditions. 
  The grey (red) bands show the significance using the default (alternative) layout \#1 (\#3). 
  The difference between the solid and hatched bands corresponds to the different pileup conditions of $\mupu=200$ and 500. 
  The band width corresponds to the significance variation due to the two models assumed for soft QCD processes.}
  \label{figure:Significance_timecut}
\end{figure}

\begin{figure}[t]
  \centering
  \includegraphics[width=1.0\columnwidth]{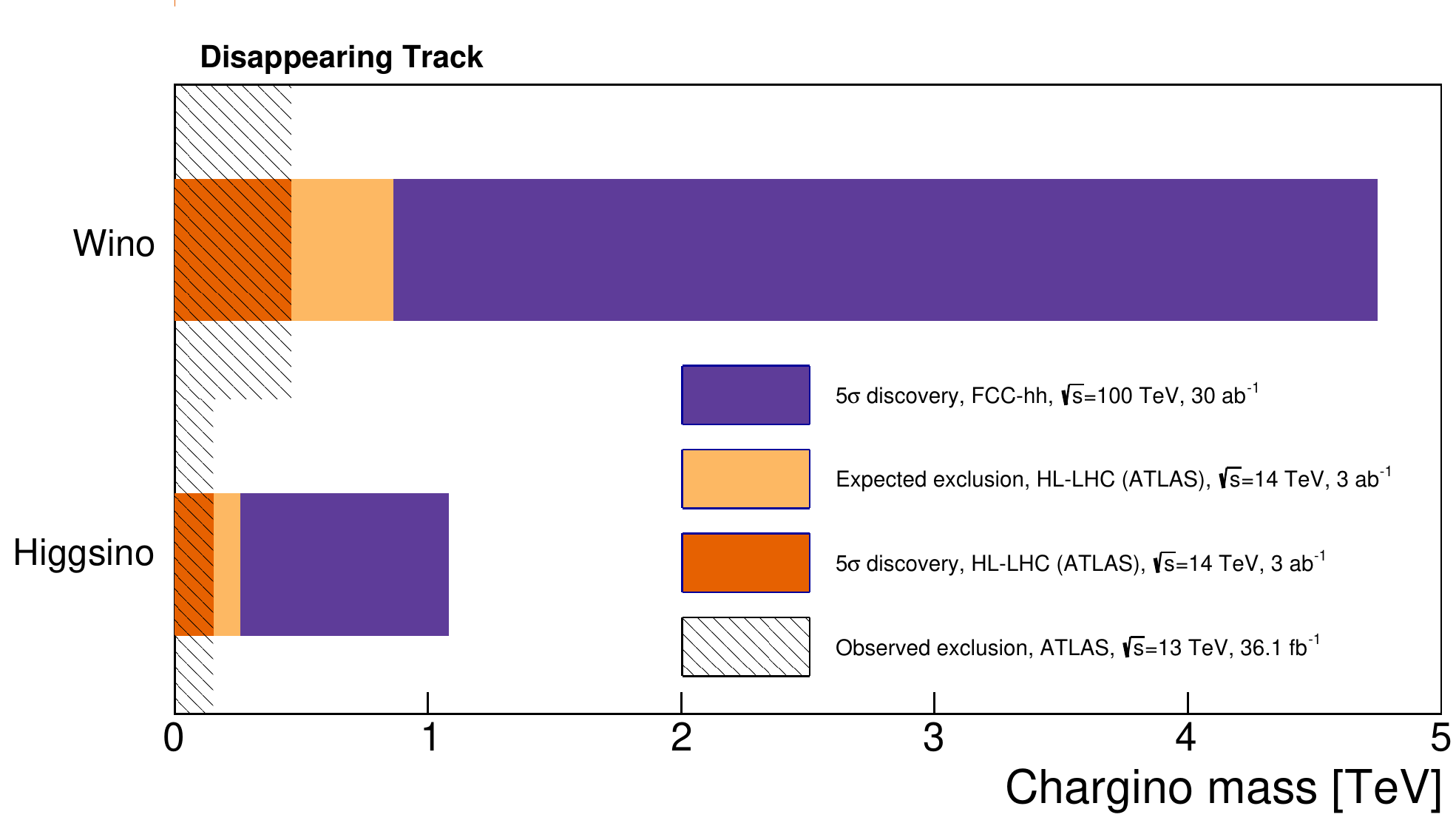}
  \caption{
   Summary of reach for charginos using a disappearing-track signature. 
   The FCC-hh 5$\sigma$-discovery reach assuming the alternative layout, the average number of $pp$ interactions per bunch crossing of 500, only non-diffractive soft-QCD processes and a use of the time information of the pixel detector is shown in blue. The reach for the wino scenario is obtained by extrapolating the sensitivity curve in Fig.~\ref{figure:Significance_timecut}.
   The 5$\sigma$-discovery reach and the 95\%~CL exclusion sensitivity are taken from~\cite{ATL-PHYS-PUB-2018-031}.
   The observed exclusion in the ATLAS search with 36~fb$^{-1}$ of Run 2 data~\cite{Aaboud:2017mpt, ATL-PHYS-PUB-2017-019} are shown as well.}
  \label{figure:summary}
\end{figure}

Figure~\ref{figure:Significance} shows the expected discovery significance for the wino~(higgsino) LSP model with the proper lifetime of 0.2~(0.023)~ns with 30 ab$^{-1}$ with the requirement of $\nhits=5$, assigning 30\% systematic uncertainty in the background yields. The background uncertainty is based on the estimate
in the ATLAS Run 2 analysis~\cite{Aaboud:2017mpt}.
The time information discussed in Sect.~\ref{subsec:fake_background} is not used in Fig.~\ref{figure:Significance}. 
The discovery significance is obtained using an approximate formula to calculate the significance of a signal+background hypothesis against a background-only hypothesis, taking into account the background uncertainty~\cite{Linnermann2003, Cranmer2005}.
The sensitivity is degraded significantly when the \mupu{} increases from 200 to 500, even if five layer hits are used in the track reconstruction.
The alternative layout (\#3) clearly improves the sensitivity, reaching well above the $5\sigma$ discovery for the 3~TeV wino in both pileup scenarios. For the 1~TeV higgsino the $5\sigma$ discovery is also feasible in both pileup scenarios but with less margin for the high pileup case.
The discovery significance is re-evaluated using the time information and the result is shown in Fig.~\ref{figure:Significance_timecut}.
The sensitivity is restored at the high pileup scenario and the degradation with increasing $\mupu${} is much more relaxed. 
The $5\sigma$ discovery is also reached with sufficient margin for the 1~TeV higgsino with 30 ab$^{-1}$ or even less if the time information is used.
The discovery sensitivities for the three tracker layouts at the pileup scenarios of $\mupu=200$ and 500 (without time information) are summarized in Table~\ref{table:ResultSensitivitymu200} and \ref{table:ResultSensitivitymu500}, respectively.
The reach of the FCC-hadron machine is shown in Fig.~\ref{figure:summary} as well as the HL-LHC reach and the current observed limits by the ATLAS experiment.
According to the present study the FCC-hadron machine has potential for answering conclusively \textit{yes} or \textit{no} to the thermal production of nearly-pure wino or higgsino dark matter.

\begin{table*}[t]\sidecaption
  \centering
  \begin{tabular}{cS[round-mode=off]S[round-mode=off]S[round-mode=off]}
    \hline\noalign{\smallskip}
    Layout        & {Default~(\#1)} & {Alternative~(\#2)} & {Alternative~(\#3)} \\
    \noalign{\smallskip}\hline\noalign{\smallskip}
                  & \multicolumn{3}{c}{wino ($m_{\tilde{\chi}_{1}^{\pm}}$ = 3~TeV)} \\
    \noalign{\smallskip}\hline\noalign{\smallskip}
    Leading jet \pT threshold [TeV] &  1  &         1    &           1   \\
    \met~threshold                [TeV] &  4  &         3    &           2   \\
    \noalign{\smallskip}\hline\noalign{\smallskip}
    Signal yield         & 28.5      & 86.5         & 287          \\
    Background  yield    & 1.9       & 7.2          & 42.6         \\
    Significance         & 10.4      & 17.8         & 26.8         \\
    \noalign{\smallskip}\hline\noalign{\smallskip}
                  & \multicolumn{3}{c}{higgsino ($m_{\tilde{\chi}_{1}^{\pm}}$ = 1~TeV)} \\
    \noalign{\smallskip}\hline\noalign{\smallskip}
    Leading jet \pT threshold [TeV] &       1 &        1     &           1  \\
    \met~threshold                [TeV] &       1 &        4     &           4  \\
    \noalign{\smallskip}\hline\noalign{\smallskip}
    Signal yield         &  2.7    & 6.6          & 19.0         \\
    Background  yield    &  673    & 1.8          & 1.6          \\
    Significance         &  0.0    & 3.4          & 8.0          \\
    \noalign{\smallskip}\hline
  \end{tabular}
  \caption{Signal and background yields as well as the discovery significance for the 3~TeV wino and 1~TeV higgsino 
  with 30 ab$^{-1}$ with the requirements of $\nhits=5$ under the pileup condition of $\mupu=200$.
 The fake-track background rate is assumed to be same for both alternative layouts. The kinematic selection requirements on 
 the leading jet \pt{} and \met{} are also given.
          }
  \label{table:ResultSensitivitymu200}
\end{table*}

\begin{table*}[t]\sidecaption
  \centering  
  \begin{tabular}{cS[round-mode=off]S[round-mode=off]S[round-mode=off]}
    \hline\noalign{\smallskip}
    Layout        & {Default~(\#1)} & {Alternative~(\#2)} & {Alternative~(\#3)} \\
    \noalign{\smallskip}\hline\noalign{\smallskip}
                  & \multicolumn{3}{c}{wino ($m_{\tilde{\chi}_{1}^{\pm}}$ = 3~TeV)} \\
    \noalign{\smallskip}\hline\noalign{\smallskip}
    Leading jet \pT threshold [TeV] &  1  &         1    &           1   \\
    \met~threshold                [TeV] &  4  &         4    &           4   \\
    \noalign{\smallskip}\hline\noalign{\smallskip}
    Signal yield         & 28.5      & 55.7         & 92.6          \\
    Background  yield    & 27.0      & 14.5         & 14.5          \\
    Significance         & 4.6       & 15.2         & 15.1           \\
    \noalign{\smallskip}\hline\noalign{\smallskip}
                  & \multicolumn{3}{c}{higgsino ($m_{\tilde{\chi}_{1}^{\pm}}$ = 1~TeV)} \\
    \noalign{\smallskip}\hline\noalign{\smallskip}
     Leading jet \pT threshold [TeV] &       1 &        2     &           8  \\
    \met~threshold                [TeV] &       1 &        6     &           8  \\
    \noalign{\smallskip}\hline\noalign{\smallskip}
    Signal yield         &  2.7       & 3.1          & 4.7         \\
    Background  yield    &  8214      & 1.6          & 0.17        \\
    Significance         &  0         & 1.8          & 4.5         \\
    \noalign{\smallskip}\hline
  \end{tabular}
  \caption{ Signal and background yields as well as the discovery significance for the 3~TeV wino and 1~TeV higgsino 
  with 30 ab$^{-1}$ with the requirements of $\nhits=5$ under the pileup condition of $\mupu=500$.
 The fake-track background rate is assumed to be same for both alternative layouts. The kinematic selection requirements on 
 the leading jet \pt{} and \met{} are also given.
          }
  \label{table:ResultSensitivitymu500}
\end{table*}

\section{Conclusions and discussions}
\label{sec:conclusion}
The discovery potential for the 3~TeV wino and 1~TeV higgsino scenarios has been evaluated using a disappearing-track signature in $pp$ collisions at $\sqrt{s}=100$~TeV with the FCC-hadron detector. 
The FCC-hh sensitivity is very promising to both the wino and higgsino scenarios, reaching up to the thermal limits, 
with inner-tracker layouts optimized for a short-track reconstruction, even in the high pileup environment of $\mupu=500$. 
If the time resolution of $\sim$50~ps is achievable at pixel-hit level, wino or higgsino dark matter will be confirmed or refuted up to the thermal limits with an integrated luminosity of 30~ab$^{-1}$ or smaller. 

Finally, we discuss several ideas that could potentially improve the sensitivity further.
The $|\eta|$ range could be expanded from $|\eta|<1$ (used in the present analysis) to e.g, $|\eta|<4$, to improve the overall signal acceptance, then split into smaller $|\eta|$ regions to optimize the sensitivity.
As shown in Fig.~\ref{figure:fakeEtaDist}, the contribution from fake tracks, that grows very rapidly with increasing $|\eta|$, will have to be evaluated for the determination of the $|\eta|$ regions. 
By tilting the sensor modules and making overlaps between them, the charginos produced at the interaction region may pass through multiple modules in a single layer. This will potentially help improving the acceptance for a short-lifetime signal without decreasing the required number of hits in the track reconstruction, therefore without increasing the fake-track background level.
It would be also beneficial if the barrel pixel layers can be moved even closer to the beam line than the alternative layout \#3. 
This has a large impact on the higgsino signal acceptance.
Another interesting avenue to pursue is to use the speed of particles relative to the speed of light ($\beta$), which can be measured using the time information of pixel hits, to reject background and characterize the observed disappearing-track signature.
With the single muon sample, the $\beta$ resolution is found to be about 14\% (with the pixel-hit time resolution of 50~ps) for the tracks with $\nhits=5$ under the alternative tracker layouts. The estimated $\beta$ resolution is roughly proportional to the pixel-hit time resolution.
The fake-track background will be reduced by requiring the measured $\beta$ to be compatible with a particle (produced from the primary $pp$ collision vertex) with a certain range of momentum and mass. Because new heavy particles move more slowly in the detector than SM particles, the $\beta$ could be used as an additional discriminant to separate the signal from scattered SM particles.

\begin{acknowledgements}
This work was supported by MEXT KAKENHI Grant number JP16K21730 and JSPS KAKENHI Grant Number JP18J11405.
\end{acknowledgements}

\bibliographystyle{spphys}       % APS-like style for physics
\bibliography{paper.bib}   % name your BibTeX data base

\begin{thebibliography}{10}
\providecommand{\url}[1]{{#1}}
\providecommand{\urlprefix}{URL }
\expandafter\ifx\csname urlstyle\endcsname\relax
  \providecommand{\doi}[1]{DOI \discretionary{}{}{}#1}\else
  \providecommand{\doi}{DOI \discretionary{}{}{}\begingroup
  \urlstyle{rm}\Url}\fi

\bibitem{HISANO200734}
J.~Hisano, S.~Matsumot, M.~Nagai, O.~Saito, M.~Senami, Phys. Lett. B
  \textbf{646}, 34  (2007).
\newblock \doi{10.1016/j.physletb.2007.01.012}.
\newblock {arXiv:hep-ph/0610249}

\bibitem{CIRELLI2007152}
M.~Cirelli, A.~Strumia, M.~Tamburini, Nucl. Phys. B \textbf{787}(1), 152
  (2007).
\newblock \doi{10.1016/j.nuclphysb.2007.07.023}.
\newblock {arXiv:0706.4071 [hep-ph]}

\bibitem{Giudice:1998xp}
G.F. Giudice, M.A. Luty, H.~Murayama, R.~Rattazzi, JHEP \textbf{12}, 027
  (1998).
\newblock \doi{10.1088/1126-6708/1998/12/027}.
\newblock {arXiv:hep-ph/9810442}

\bibitem{Randall:1998uk}
L.~Randall, R.~Sundrum, Nucl. Phys. B \textbf{557}, 79 (1999).
\newblock \doi{10.1016/S0550-3213(99)00359-4}.
\newblock {arXiv:hep-th/9810155}

\bibitem{IBE2012374}
M.~Ibe, T.T. Yanagida, Phys. Lett. B \textbf{709}(4), 374 (2012).
\newblock \doi{https://doi.org/10.1016/j.physletb.2012.02.034}.
\newblock {arXiv:1112.2462 [hep-ph]}

\bibitem{PhysRevD.85.095011}
M.~Ibe, S.~Matsumoto, T.T. Yanagida, Phys. Rev. D \textbf{85}, 095011 (2012).
\newblock \doi{10.1103/PhysRevD.85.095011}.
\newblock {arXiv:1202.2253 [hep-ph]}

\bibitem{PhysRevD.87.015028}
B.~Bhattacherjee, B.~Feldstein, M.~Ibe, S.~Matsumoto, T.T. Yanagida, Phys. Rev.
  D \textbf{87}, 015028 (2013).
\newblock \doi{10.1103/PhysRevD.87.015028}.
\newblock {arXiv:1207.5453 [hep-ph]}

\bibitem{Cohen_2013}
T.~Cohen, M.~Lisanti, A.~Pierce, T.R. Slatyer, J. Cosmol. Astropart. Phys.
  \textbf{2013}(10), 61 (2013).
\newblock \doi{10.1088/1475-7516/2013/10/061}.
\newblock {arXiv:1307.4082 [hep-ph]}

\bibitem{Fan2013}
J.~Fan, M.~Reece, JHEP \textbf{10}, 124 (2013).
\newblock \doi{10.1007/JHEP10(2013)124}.
\newblock {arXiv:1307.4400 [hep-ph]}

\bibitem{PhysRevD.94.115019}
H.~Baer, V.~Barger, H.~Serce, Phys. Rev. D \textbf{94}, 115019 (2016).
\newblock \doi{10.1103/PhysRevD.94.115019}.
\newblock {arXiv:1609.06735 [hep-ph]}

\bibitem{Nesti_2013}
F.~Nesti, P.~Salucci, J. Cosmol. Astropart. Phys. \textbf{2013}(07), 16 (2013).
\newblock \doi{10.1088/1475-7516/2013/07/016}.
\newblock {arXiv:1304.5127 [astro-ph.GA]}

\bibitem{Sugai2016}
H.~Sugai, K.~Hayashi, K.~Ichikawa, M.N. Ishigaki, S.~Matsumoto, M.~Ibe, MNRAS
  \textbf{461}(3), 2914 (2016).
\newblock \doi{10.1093/mnras/stw1457}.
\newblock {arXiv:1603.08046 [astro-ph.GA]}

\bibitem{HISANO2010311}
J.~Hisano, K.~Ishiwata, N.~Nagata, Phys. Lett. B \textbf{690}(3), 311 (2010).
\newblock \doi{https://doi.org/10.1016/j.physletb.2010.05.047}.
\newblock {arXiv:1004.4090 [hep-ph]}

\bibitem{Nagata2015}
N.~Nagata, S.~Shirai, {JHEP} \textbf{2015}, 29 (2015).
\newblock \doi{10.1007/JHEP01(2015)029}.
\newblock {arXiv:1410.4549 [hep-ph]}

\bibitem{Baer2018}
H.~Baer, V.~Barger, D.~Sengupta, X.~Tata, Eur. Phys. J. C \textbf{78}(10), 838
  (2018).
\newblock \doi{10.1140/epjc/s10052-018-6306-y}.
\newblock {arXiv:1803.11210 [hep-ph]}

\bibitem{ARKANIHAMED2006108}
N.~Arkani-Hamed, A.~Delgado, G.~Giudice, Nucl. Phys. B \textbf{741}(1), 108
  (2006).
\newblock \doi{https://doi.org/10.1016/j.nuclphysb.2006.02.010}.
\newblock {arXiv:hep-ph/0601041}

\bibitem{BADZIAK2017226}
M.~Badziak, M.~Olechowski, P.~Szczerbiak, Phys. Lett. B \textbf{770}, 226
  (2017).
\newblock \doi{https://doi.org/10.1016/j.physletb.2017.04.059}.
\newblock {arXiv:1701.05869 [hep-ph]}

\bibitem{Beneke2017}
M.~Beneke, A.~Bharucha, A.~Hryczuk, S.~Recksiegel, P.~Ruiz-Femen{\'i}a, JHEP
  \textbf{01}, 002 (2017).
\newblock \doi{10.1007/JHEP01(2017)002}.
\newblock {arXiv:1611.00804 [hep-ph]}

\bibitem{Kowalska2018}
K.~Kowalska, E.M. Sessolo, Adv. High Energy Phys. \textbf{2018}(6828560), 15
  (2018).
\newblock \doi{10.1155/2018/6828560}.
\newblock {arXiv:1802.04097 [hep-ph]}

\bibitem{Ibe:2012sx}
M.~Ibe, S.~Matsumoto, R.~Sato, Phys. Lett. B \textbf{721}, 252 (2013).
\newblock \doi{10.1016/j.physletb.2013.03.015}.
\newblock {arXiv:1212.5989 [hep-ph]}

\bibitem{1126-6708-2003-03-045}
A.J. Barr, C.G. Lester, M.A. Parker, B.C. Allanach, P.~Richardson, {JHEP}
  \textbf{03}, 045 (2003).
\newblock \doi{10.1088/1126-6708/2003/03/045}.
\newblock {arXiv:hep-ph/0208214}

\bibitem{PhysRevD.55.330}
{C.-H. Chen}, M.~Drees, J.~F.~Gunion, Phys. Rev. D \textbf{55}, 330 (1997).
\newblock \doi{10.1103/PhysRevD.55.330}.
\newblock {arXiv:hep-ph/9607421}

\bibitem{Chen:1999yf}
{C.-H. Chen}, M.~Drees, J.~F.~Gunion,   (1999).
\newblock {arXiv:hep-ph/9902309}

\bibitem{Asai2008185}
S.~Asai, T.~Moroi, T.~T.~Yanagida, Phys. Lett. B \textbf{664}, 185 (2008).
\newblock \doi{10.1016/j.physletb.2008.05.019}.
\newblock {arXiv:0802.3725 [hep-ph]}

\bibitem{Low2014}
M.~Low, {L.-T. Wang}, {JHEP} \textbf{08}, 161 (2014).
\newblock \doi{10.1007/JHEP08(2014)161}.
\newblock {arXiv:1404.0682 [hep-ph]}

\bibitem{Cirelli2015}
M.~Cirelli, F.~Sala, M.~Taoso, {JHEP} \textbf{10}, 033 (2014).
\newblock \doi{10.1007/JHEP10(2014)033}.
\newblock {arXiv:1407.7058 [hep-ph]. Erratum: JHEP01(2015)041}

\bibitem{Mahbubani2017}
R.~Mahbubani, P.~Schwaller, J.~Zurita, {JHEP} \textbf{06}, 119 (2017).
\newblock \doi{10.1007/JHEP06(2017)119}.
\newblock {arXiv:1703.05327 [hep-ph]}

\bibitem{FUKUDA2018306}
H.~Fukuda, N.~Nagata, H.~Otono, S.~Shirai, Phys. Lett. B \textbf{781}, 306
  (2018).
\newblock \doi{10.1016/j.physletb.2018.03.088}.
\newblock {arXiv:1703.09675 [hep-ph]}

\bibitem{PhysRevD.98.035026}
T.~Han, S.~Mukhopadhyay, X.~Wang, Phys. Rev. D \textbf{98}, 035026 (2018).
\newblock \doi{10.1103/PhysRevD.98.035026}.
\newblock {arXiv:1805.00015 [hep-ph]}

\bibitem{Cao2018}
Q.H. Cao, T.~Gong, K.P. Xie, Z.~Zhen.
\newblock {Measuring Relic Abundance of Minimal Dark Matter at Hadron
  Colliders} (2018).
\newblock {arXiv:1810.07658 [hep-ph]}

\bibitem{Allanach:2001kg}
B.C. Allanach, Comput. Phys. Commun. \textbf{143}, 305 (2002).
\newblock \doi{10.1016/S0010-4655(01)00460-X}.
\newblock {arXiv:hep-ph/0104145}

\bibitem{Djouadi:2006bz}
A.~Djouadi, M.M. Muhlleitner, M.~Spira, Acta Phys. Polon. \textbf{B38}, 635
  (2007).
\newblock {arXiv:hep-ph/0609292}

\bibitem{Beenakker:1996ch}
W.~Beenakker, R.~H{\"o}pker, M.~Spira, P.~Zerwas, Nucl. Phys. B \textbf{492},
  51 (1997).
\newblock \doi{10.1016/S0550-3213(97)00084-9}.
\newblock {arXiv:hep-ph/9610490}

\bibitem{Alwall:2014}
J.~Alwall, R.~Frederix, S.~Frixione, V.~Hirschi, F.~Maltoni, O.~Mattelaer, H.S.
  Shao, T.~Stelzer, P.~Torrielli, M.~Zaro, JHEP \textbf{07}, 079 (2014).
\newblock \doi{10.1007/JHEP07(2014)079}.
\newblock {arXiv:1405.0301 [hep-ph]}

\bibitem{Sjostrand:2014zea}
T.~Sj{\"o}strand, S.~Ask, J.R. Christiansen, R.~Corke, N.~Desai, P.~Ilten,
  S.~Mrenna, S.~Prestel, C.O. Rasmussen, P.Z. Skands, Comput. Phys. Commun.
  \textbf{191}, 159 (2015).
\newblock \doi{10.1016/j.cpc.2015.01.024}.
\newblock {arXiv:1410.3012 [hep-ph]}

\bibitem{deFavereau2014}
J.~de~Favereau, C.~Delaere, P.~Demin, A.~Giammanco, V.~Lemaitre, A.~Mertens,
  M.~Selvaggi, JHEP \textbf{02}, 057 (2014).
\newblock \doi{10.1007/JHEP02(2014)057}.
\newblock {arXiv:1307.6346 [hep-ex]}

\bibitem{Cacciari:2008gp}
M.~Cacciari, G.P. Salam, G.~Soyez, JHEP \textbf{04}, 063 (2008).
\newblock \doi{10.1088/1126-6708/2008/04/063}.
\newblock {arXiv:0802.1189 [hep-ph]}

\bibitem{FCC-hh_detector}
{M. Benedikt et al}, {Future Circular Collider Study. Volume 3: The Hadron
  Collider (FCC-hh) Conceptual Design Report}.
\newblock {CERN-ACC-2018-0058} (2018).
\newblock {Submitted to Eur. Phys. J. ST.}

\bibitem{Aaboud:2017mpt}
{M.~Aaboud et al., [ATLAS Collaboration]}, JHEP \textbf{06}, 022 (2018).
\newblock \doi{10.1007/JHEP06(2018)022}.
\newblock {arXiv:1712.02118 [hep-ex]}

\bibitem{ATLASMaterialBudget}
{M.~Aaboud et al., [ATLAS Collaboration]}, JINST \textbf{12}, P12009 (2017).
\newblock {arXiv:1707.02826 [hep-ex]}

\bibitem{Agostinelli:2002hh}
S.~Agostinelli, et~al., Nucl. Instrum. Meth. A \textbf{506}, 250 (2003).
\newblock \doi{10.1016/S0168-9002(03)01368-8}

\bibitem{Pellegrini2014}
G.~Pellegrini, P.~Fern\'{a}ndez-Martínez, M.~Baselga, C.~Fleta, D.~Flores,
  V.~Greco, S.~Hidalgo, I.~Mandi\'{c}, G.~Kramberger, D.~Quirion, M.~Ullan,
  Nucl. Instrum. Meth. A \textbf{765}, 12 (2014).
\newblock \doi{10.1016/j.nima.2014.06.008}

\bibitem{Lange2017}
J.~Lange, M.~Carulla, E.~Cavallaro, L.~Chytka, P.~Davis, D.~Flores,
  F.~F\"{o}rster, S.~Grinstein, S.~Hidalgo, T.~Komarek, G.~Kramberger,
  I.~Mandi\'{c}, A.~Merlos, L.~Nozka, G.~Pellegrini, D.~Quirion, T.~Sykora,
  JINST \textbf{12}, P05003 (2017).
\newblock \doi{10.1088/1748-0221/12/05/P05003}.
\newblock {arXiv:1703.09004 [physics.ins-det]}

\bibitem{Linnermann2003}
J.~Linnemann.
\newblock {Measures of Significance in HEP and Astrophysics} (2003).
\newblock {arXiv:physics/0312059}

\bibitem{Cranmer2005}
K.~Cranmer.
\newblock {Statistical Challenges for Searches for New Physics at the LHC}
  (2005).
\newblock {arXiv:physics/0511028}

\bibitem{ATL-PHYS-PUB-2018-031}
{ATLAS Collaboration}.
\newblock {ATLAS sensitivity to winos and higgsinos with a highly compressed
  mass spectrum at the HL-LHC}.
\newblock {ATL-PHYS-PUB-2018-031} (2018).
\newblock {http://cds.cern.ch/record/2647294}

\bibitem{ATL-PHYS-PUB-2017-019}
{ATLAS Collaboration}.
\newblock {Search for direct pair production of higgsinos by the
  reinterpretation of the disappearing track analysis with 36.1 fb${}^{-1}$ of
  $\sqrt{s} = 13\;\mbox{TeV}$ data collected with the ATLAS exepriment}.
\newblock {ATL-PHYS-PUB-2017-019} (2017).
\newblock {http://cdsweb.cern.ch/record/2297480}

\end{thebibliography}

\end{document}